\documentclass[aps,prd,preprint,preprintnumbers,unsortedaddress,superscriptaddress,showpacs,nofootinbib]{revtex4-1}
\usepackage{graphicx,color}
\usepackage{relsize}
\usepackage{slashed}
\usepackage{color}
\usepackage{ulem}
\usepackage{tabu}
\usepackage{float}

\usepackage{amsmath}
\usepackage{amssymb}
\usepackage{amsthm}
\usepackage{mathrsfs}
\usepackage{graphicx}
\usepackage{epstopdf}
\usepackage{fancyhdr}
\usepackage{array}
\usepackage[all]{xy}
\usepackage{eufrak}
\usepackage{euscript}
\usepackage{enumerate}
\usepackage{slashed}
\usepackage{hyperref}
\usepackage{subfigure} % NEEDED IN ORDER TO USE SUBFIG
\usepackage{epstopdf} % COMPILE EPS (AND PDF) FIGURES USING PDFLATEX!!!

\usepackage{hyperref}
\hypersetup{pdftex,colorlinks=true,linkcolor=blue,citecolor=blue,menucolor=black,urlcolor=blue,filecolor=blue}

\newcommand{\myvec}[1]{\ensuremath{\begin{pmatrix}#1\end{pmatrix}}}
\begin{document}

\title{Doubly charmed $\Xi_{cc}$ molecular states from meson-baryon interaction}

\author{J.~M.~Dias}
\email{jdias@if.usp.br}
\affiliation{Institute of Modern Physics, Chinese Academy of Sciences, Lanzhou 730000, China
}
 
\affiliation{Departamento de F\'isica Te\'orica and IFIC, Centro Mixto Universidad de Valencia-CSIC, 
Institutos de Investigaci\'on de Paterna, Aptdo. 22085, 46071 Valencia, Spain.
}
\affiliation{
Instituto de F\'isica, Universidade de S\~ao Paulo, C.P. 66318, 05389-970 S\~ao 
Paulo, SP, Brazil.
}

\author{V.~R.~Debastiani}
\email{vinicius.rodrigues@ific.uv.es}
\affiliation{Institute of Modern Physics, Chinese Academy of Sciences, Lanzhou 730000, China
}
\affiliation{Departamento de F\'isica Te\'orica and IFIC, Centro Mixto Universidad de Valencia-CSIC, 
Institutos de Investigaci\'on de Paterna, Aptdo. 22085, 46071 Valencia, Spain.
}
\author{Ju-Jun Xie}
\email{xiejujun@impcas.ac.cn}
\affiliation{Institute of Modern Physics, Chinese Academy of Sciences, Lanzhou 730000, China
}

\author{E.~Oset}
\email{eulogio.oset@ific.uv.es}
\affiliation{Institute of Modern Physics, Chinese Academy of Sciences, Lanzhou 730000, China
}
\affiliation{Departamento de F\'isica Te\'orica and IFIC, Centro Mixto Universidad de Valencia-CSIC, 
Institutos de Investigaci\'on de Paterna, Aptdo. 22085, 46071 Valencia, Spain.
}

\preprint{}

%\date{\today}

\begin{abstract}
Stimulated by the new experimental LHCb findings associated 
with the $\Omega_c$ states, some of which we have 
described in a previous work as being dynamically generated through meson-baryon 
interaction, we have extended this approach to make predictions 
for new $\Xi_{cc}$ molecular states in the $C=2$, $S=0$ and 
$I=1/2$ sector. These states manifest themselves as 
poles in the solution of the Bethe-Salpeter equation in coupled 
channels. The kernels of this equation were obtained using the 
Lagrangians coming from the hidden local gauge symmetry, where 
the interactions are dominated by the exchange of light vector mesons. 
The extension of this approach to the heavy sector stems from the 
realization that the dominant interaction corresponds to having the 
heavy quarks as spectators, which implies the preservation of 
the heavy quark symmetry. As a result, we get several states: two 
states from the pseudoscalar meson-baryon interaction with $J^P=1/2^-$, 
and masses around $4080$ and $4090$ MeV, and one at $4150$ MeV for 
$J^P=3/2^-$. Furthermore, from the vector meson-baryon interaction 
we get three states degenerate with $J^P=1/2^-$ and $3/2^-$ from $4220$ 
MeV to $4330$ MeV, and two more states around $4280$ MeV and $4410$ MeV, 
degenerate with $J^P=1/2^-,\, 3/2^-$ and $5/2^-$.
\end{abstract}

\pacs{14.40.Rt, 12.40.Yx, 13.75.Lb}

\maketitle

%%%%%%%%%%%%%%%%%%%%%%%%%%%%%%%%%%%%%%%%%%%%%%%%%%%%%%%%%%%%%%%%%%%%%%%%%%%%%%%%%%% 
\section{Introduction}
%%%%%%%%%%%%%%%%%%%%%%%%%%%%%%%%%%%%%%%%%%%%%%%%%%%%%%%%%%%%%%%%%%%%%%%%%%%%%%%%%%%

Over the last decade, the field of hadron spectroscopy is living a new era due to 
a large bulk of experimental results, which has triggered an intense theoretical 
activity in order to describe and understand these experimental data. They are 
challenging our knowledge of hadron dynamics since many states 
cannot be accommodated within the standard picture for the hadron. In 2015, the 
LHCb reported the observation of the states $P_c^+(4380)$ and $P_c^+(4450)$ 
in the $J/\psi p$ invariant mass distribution \cite{penta1,penta2,penta3} and, afterwards, 
five narrow $\Omega_c$ states \cite{omega} were measured in the $\Xi_c^+K^-$ 
mass spectrum. Especially interesting was the observation of a doubly 
charmed baryon (DCB), called $\Xi_{cc}^{++}$, recently seen by the LHCb collaboration in the 
$\Lambda_c^+ K^- \pi^+ \pi^+$ final state, with mass around $3621$ MeV 
\cite{lhcbXicc}. This value is higher than that for the first doubly 
charmed state $\Xi_{cc}^{+}$ measured in the $\Lambda_c^+K^-\pi^+$ 
mass spectrum, by SELEX in 2002 \cite{selex1}, and later confirmed in Ref.~\cite{selex2} by 
the same collaboration. However, this latter state was not confirmed by 
FOCUS \cite{focus}, Belle \cite{belle}, BABAR \cite{babar} and the LHCb \cite{lhcb} 
collaborations.

On the theoretical side, a DCB state with a mass similar to that one 
reported by the LHCb had been predicted in Ref.~\cite{rujula}, using a renormalizable 
gauge field theory. A DCB was also predicted in Ref.~\cite{ebert}, 
where the relativistic quark-diquark potential model was employed. Using the one gluon 
exchange model, the authors of Ref.~\cite{gluexch} had also predicted a 
doubly heavy baryon state in which the mass value obtained is close to the one 
measured by the LHCb. In particular, these works advocate that the $\Xi_{cc}^{++}$ should 
be accommodated in the quark picture. On the other hand, many other theoretical 
approaches were used to study doubly charmed baryon states \footnote{We refer the reader to Ref.~\cite{slz}, which presents a review of the literature on those works.} before 
the LHCb measurements, including even triply heavy baryons extended to the beauty sector 
\cite{jaffe}.

More recently, in particular after the LHCb announcement of the newly 
$\Xi^{++}_{cc}$, a new wave of theoretical studies have aroused 
in an attempt to understand its properties, including also new predictions. 
In Ref.~\cite{magmon}, the chiral corrections were employed to estimate the 
magnetic moments of DCB with $J=1/2$. Weak decays were studied 
in Refs.~\cite{weak,weak2}, strong and radiative decays were investigated 
in Ref.~\cite{radia}, and QCD sum rules were used in Ref.~\cite{qcdsr}. The 
molecular picture was also adopted. In Ref.~\cite{harada}, 
the meson-baryon transitions between the coupled channels 
$J/\psi N-\Lambda_c\bar{D}^{(*)}-\Sigma_c^{(*)}\bar{D}^{(*)}$ were constructed 
taking into account the pion and $D^{(*)}$ meson exchange, and then used as 
the potential in the complex scaled Schr{\"o}dinger-type 
equation. According to their findings, if the $P_c(4380)$ exists as a hadronic 
molecular state, the existence of a $\Xi^*_{cc}(4380)$ with 
almost the same mass as the $P_c(4380)$ state should be expected. 
Studying the same type of interaction, that is, the 
meson-baryon one, in Ref.~\cite{zguo} an $S$-wave scattering of ground state 
doubly charmed baryons ($\Xi^{++}_{cc},\,\Xi^{+}_{cc},\,\Omega^+_{cc}$) and 
the light pseudoscalar ($\pi,\, K,\, \eta$) mesons was implemented by 
means of chiral effective theory, and several DCB 
resonances were predicted. This is particularly interesting and, since 
the LHCb has observed $\Omega_c$ resonances, we can expect 
there may exist $\Xi_{cc}$ resonances as well. In view of this, 
in this work we study the meson-baryon interaction in order to 
investigate DCB resonances that can be 
confronted with the experimental measurements to be made in the 
near future. 

Among the models employed to study meson-baryon interactions, one 
is particularly interesting and powerful to describe the 
meson-baryon interaction. It combines chiral dynamics with unitarity 
in coupled channels, named as chiral unitary approach. Using 
chiral Lagrangians we can obtain the transition amplitudes between all 
the relevant channels contributing to the interaction we are concerned. 
Then, these amplitudes are unitarized through the Bethe-Salpeter 
equation, from which bound states/resonances emerge as solutions 
in the complex energy plane. We say that these bound states/resonances 
are dynamically generated. One famous example was the long-standing 
two $\Lambda(1405)$ states \cite{lamb1,lamb2,lamb3,lamb4}. 
%In Ref.~\cite{lamb1} it was described as dynamically generated through the 
%pseudoscalar meson-baryon interaction ($PB$). More concretely, assuming 
%coupled channels, mostly the $\bar{K}N$ and $\pi \Sigma$ channels, it was 
%found in Refs.~\cite{lamb1,lamb3} that a double pole structure appears close to the 
%$\Lambda(1405)$ mass region. 

The extension of the chiral unitary approach to describe vector meson-baryon 
interactions was done in Refs.~\cite{hgmb1,hgmb2}, where the authors used the 
Lagrangians from the hidden gauge approach \cite{hga1,hga2,hga3}, 
which extends the chiral Lagrangians to include vector mesons. Its extension 
to the charm sector was done in Refs.~\cite{,vbmixpb1,vbmixpb2,lutz,mlzu}. 
In particular, in Refs.~\cite{vbmixpb1,vbmixpb2} it was found that in the dominant 
terms of the interaction the heavy quarks are spectators and, hence, the 
dominant contributions come from the exchange of light vector mesons. As a 
consequence, this approach satisfies the heavy quark spin symmetry, which 
is the symmetry of QCD that says the interaction is independent of the spin of 
the heavy quarks in the limit of the heavy quark mass going to infinity.

In Ref.~\cite{ourpaper}, we have done such an extension of the 
hidden gauge to the charm sector taking into account the spin-flavor 
wave function for baryons and, since the heavy quarks act as spectators, 
the vector-baryon-baryon ($VBB$) vertex for the diagonal terms was obtained using the $SU(3)$ 
content of $SU(4)$. 
The work shares many elements and similar results with the work of 
Ref.~\cite{montana}, where baryon wave functions within $SU(4)$ are 
strictly used. A different approach motivated by the works of 
Refs.~\cite{montana,ourpaper,romanets} is done in Ref.~\cite{pavao}, where some 
states obtained are also associated with the observed states. In particular, 
this was done after reviewing the renormalization scheme of 
Ref.~\cite{romanets}, where some $\Omega_c$ states had been predicted 
with masses smaller than the ones observed by the LHCb. 

Following the approach of Ref.~\cite{ourpaper}, we have described 
some of the $\Omega_c$ states observed by the LHCb collaboration. A remarkable 
agreement with the experimental results for those singly charmed baryon 
states were obtained. For $J^P=1/2^-$ two states can be 
related to the observed ones, the $\Omega_c(3090)$ and $\Omega_c(3050)$. 
In addition, another pole with $J^P=3/2^-$ could be related to the $\Omega_c(3119)$. 
Motived by this remarkable agreement, we have also employed the same approach, 
but this time to predict singly heavy baryon resonances in the beauty 
sector \cite{omegab}, named as $\Omega_b$ states. In view of this, and 
stimulated by the LHCb recent discovery of a doubly charmed 
baryon structure, $\Xi^{++}_{cc}$, we have used this same approach in order to investigate 
DCB states, that can be dynamically generated through the 
interaction between doubly and singly charmed baryon with pseudoscalar 
and vector mesons with or without charm. 

The planned experiments, for instance, like the one at the FAIR facility 
will involve studies of charm physics, and 
the observation of such new states, certainly will shed light on the debate 
about their quantum numbers, production mechanism and quark content. 
This will certainly be a good scenario to test most of the models 
which are used to understand those states from the multiquark point of view.

%%%%%%%%%%%%%%%%%%%%%%%%%%%%%%%%%%%%%%%%%%%%%%%%%%%%%%%%%%%%%%%%%%%%%%%%%%%%%%%%%%% 
\section{Theoretical Framework}
%%%%%%%%%%%%%%%%%%%%%%%%%%%%%%%%%%%%%%%%%%%%%%%%%%%%%%%%%%%%%%%%%%%%%%%%%%%%%%%%%%%

In order to obtain the transition matrix elements using the Bethe-Salpeter 
equation, we must write down the relevant space of states
in the $C=2,\,S=0$ and $I=1/2$ sector, which are the channels 
contributing to the meson-baryon interaction in $S$-wave we are concerned with. 
We use the channels established in Ref.~\cite{romanets} and separate them into 
different cases from the interaction of baryons ($J^P=1/2^+,\, 3/2^+$) 
with pseudoscalar ($0^-$) and vector mesons ($1^-$), 
as it was done in Ref.~\cite{ourpaper} for the case of the $\Omega_c$ states. We 
should emphasize here that the channels listed in Ref.~\cite{romanets} were written 
taking into account the previous value for the $\Xi_{cc}$ mass which was $3519$ 
MeV. In this work we are considering the same channels of Ref.~\cite{romanets} but using the 
value reported by the LHCb collaboration \cite{lhcbXicc}, which is equal to $3621$ MeV. 
This means we are considering this new value as the ground state for the $\Xi_{cc}$ baryon. 
Accordingly, we also update the estimate for the excited $\Xi^*_{cc}$, 
taking its mass as $81$ MeV higher than that of $\Xi_{cc}$, similar to the 
$\Xi^{\prime}_c-\Xi^*_c$ mass splitting, as done in Ref.~\cite{romanets}. The estimates 
for the $\Omega_{cc}$ and $\Omega^*_{cc}$ masses are also taken as the same adopted 
in Ref.~\cite{romanets}, given in Ref.~\cite{juan}. The other masses are taken as 
isospin averages of the ones listed by the Particle Data Group \cite{pdg}. In 
Tables \ref{tab1}, \ref{tab1new}, \ref{tab2} and \ref{tab2new} we show the channels and their respective 
thresholds reevaluated taking $M_{\Xi_{cc}}=3621$ MeV.
\begin{table}[h!]
\caption{Baryon-pseudoscalar states ($J^P=1/2^-$) chosen and threshold mass in MeV.}
\centering
\begin{tabular}{c | c c c c c c c}
\hline\hline
{\bf Channel} ~& ~$\Xi_{cc}\pi$~ & ~$\Lambda_c D$~ & ~$\Xi_{cc} \eta$~ & ~$\Omega_{cc} K$ ~&~ $\Sigma_c D$ ~& ~$\Xi_c D_s$ ~& ~$\Xi^{\prime}_c D_s$\\
\hline
{\bf Threshold} ~& $3759$ & $4154$ & $4169$ & $4208$ & $4321$ & $4438$ & $4545$\\
\hline\hline
\end{tabular}
\label{tab1}
\end{table}

\begin{table}[h!]
\caption{Baryon-pseudoscalar states ($J^P=3/2^-$) chosen and threshold mass in MeV.}
\centering
\begin{tabular}{c | c c c c c}
\hline\hline
{\bf Channel} ~& ~$\Xi^*_{cc}\pi$~ & ~$\Xi^*_{cc} \eta$~ & ~$\Omega^*_{cc} K$ ~&~ $\Sigma^*_c D$ ~& ~$\Xi^*_c D_s$\\
\hline
{\bf Threshold} ~& $3840$ & $4250$ & $4291$ & $4385$ & $4615$\\
\hline\hline
\end{tabular}
\label{tab1new}
\end{table}

\begin{table}[h!]
\caption{Baryon-vector meson states ($J^P=1/2^-,\,3/2^-$) chosen and threshold mass in MeV.}
\centering
\begin{tabular}{c | c c c c c c c c}
\hline\hline
{\bf Channel} ~& ~$\Lambda_c D^*$~ & ~$\Xi_{cc} \rho$~ & ~$\Xi_{cc} \omega$~ & ~$\Sigma_c D^*$ ~& ~$\Xi_c D^*_s$ ~& ~$\Omega_{cc} K^*$~&~ $\Xi_{cc} \phi$  ~&~ $\Xi_c^{\prime} D_s^*$\\
\hline
{\bf Threshold} ~& $4295$ & $4397$ & $4404$ & $4462$ & $4582$ & $4606$ & $4641$ & $4689$\\
\hline\hline
\end{tabular}
\label{tab2}
\end{table}

\begin{table}[h!]
\caption{Baryon-vector meson states ($J^P=1/2^-,\,3/2^-,\,5/2^-$) chosen and threshold mass in MeV.}
\centering
\begin{tabular}{c | c c c c c c}
\hline\hline
{\bf Channel} ~& ~$\Xi^*_{cc} \rho$~ & ~$\Xi^*_{cc} \omega$~ & ~$\Sigma^*_c D^*$ ~&~$\Omega^*_{cc} K^*$ ~& ~$\Xi^*_{cc} \phi$ ~&~ $\Xi^*_c D_s^*$\\
\hline
{\bf Threshold} ~& $4478$ & $4485$ & $4526$ & $4689$ & $4722$ & $4759$\\
\hline\hline
\end{tabular}
\label{tab2new}
\end{table}

Next, we will discuss the use of Lagrangians from hidden local gauge symmetry which provide an 
easy manner to evaluate the meson-baryon interaction involving the channels listed in Tables \ref{tab1}, \ref{tab1new}, \ref{tab2} and \ref{tab2new}.

\subsection{Transition amplitudes}

The use of chiral Lagrangians to calculate the transition amplitudes is complicated 
when states in the charm sector are involved. This happens because 
one needs to extend those Lagrangians from $SU(3)$ to $SU(4)$ and the use of this latter 
symmetry must be handled with care when dealing with mesons and baryons with 
such disparate masses. On the other hand, the use 
of the Lagrangians coming from the hidden local gauge symmetry allows us to make use of the 
$SU(3)$ content of $SU(4)$ since the heavy quark is treated as a spectator in our formalism. 
As a consequence the rules of heavy quark spin symmetry are fulfilled \cite{liang} 
for the dominant diagonal interactions.

\begin{figure}
	\begin{center} 
		\includegraphics[width=0.4\textwidth]{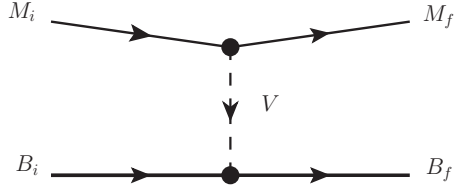}
	\end{center} 
	\caption{\label{MBdiag}Diagram representing the meson-baryon interaction through vector 
	meson exchange. $M_i(M_f)$ and $B_i(B_f)$ are the initial (final) meson and baryon states, 
	respectively, taking place on the interaction, while $V$ stands for the vector meson exchanged. }
\end{figure}

In the local hidden gauge approach in $SU(3)$, the meson-baryon interaction 
proceeds by means of vector meson exchange as illustrated in 
Fig.~\ref{MBdiag}. According to the hidden local gauge approach, the vector-pseudoscalar-pseudoscalar
coupling ($VPP$), i. e. the upper vertex of the diagram depicted in Fig.~\ref{MBdiag} is described 
by the following Lagrangian
\begin{equation}\label{vpplag}
\mathcal{L}_{VPP}=-ig \langle\,\, [\phi, \partial_{\mu}\phi] V^{\mu} \rangle \,  ,
\end{equation}
where $\phi$ and $V^{\mu}$ are the $SU(3)$ matrices for pseudoscalar and vector mesons, respectively, 
given by
\begin{equation}
\label{phimatrix}
\phi =
\left(
\begin{array}{ccc}
\frac{1}{\sqrt{2}} \pi^0 + \frac{1}{\sqrt{6}} \eta & \pi^+ & K^+ \\
\pi^- & - \frac{1}{\sqrt{2}} \pi^0 + \frac{1}{\sqrt{6}} \eta & K^0 \\
K^- & \bar{K}^0 & - \frac{2}{\sqrt{6}} \eta
\end{array}
\right)\, ,
\end{equation}
and
\begin{equation}
\label{vfields}
V_\mu=\left(
\begin{array}{ccc}
\frac{\rho^0}{\sqrt{2}}+\frac{\omega}{\sqrt{2}} & \rho^+ & K^{*+}\\
\rho^- &-\frac{\rho^0}{\sqrt{2}}+\frac{\omega}{\sqrt{2}} & K^{*0}\\
K^{*-} & \bar{K}^{*0} &\phi\\
\end{array}
\right)_\mu\ ,
\end{equation}
while the symbol $\langle \, . \, .\, . \rangle$ in Eq.~\eqref{vpplag} stands for the 
$SU(3)$ trace and the coupling $g=M_V/2f_{\pi}$, with $f_{\pi}=93$ MeV being the pion decay constant. 
The extension of Eq.~\eqref{vpplag} to $SU(4)$ is straightforward and the discussion on 
how to do this can be found in Refs.~\cite{vbmixpb1,vbmixpb2}. Alternatively one can 
use an explicit method, rather clarifying in the case of exchange of light vectors mesons, 
which leaves the heavy quarks as spectators, consisting in explicitly writing the 
operators for vector exchange in terms of quarks. This is done in Ref.~\cite{roca} 
and the conclusion is that one can use directly Eq.~\eqref{vpplag} with $SU(4)$ 
matrices for $\phi$ and $V_{\mu}$ \cite{vbmixpb1,vbmixpb2} and the procedure projects 
automatically in $SU(3)$ in the case that the heavy quarks are spectators.

On the other hand, the vector-baryon-baryon ($VBB$) in $SU(3)$, can 
be calculated within the local hidden gauge formalism using the following Lagrangian,
\begin{equation}\label{vbblag}
\mathcal{L}_{VBB}=g \Big( \langle \bar{B}\gamma_{\mu}  [V^{\mu},B]\, \rangle 
+ \langle \bar{B}\gamma_{\mu} B \rangle \langle V^{\mu} \rangle \Big)\, ,
\end{equation}
with $B$ associated with the $SU(3)$ matrix for the baryon octet, which is 
\begin{equation}
\label{Bmatrix}
B =
\left(
\begin{array}{ccc}
\frac{1}{\sqrt{2}} \Sigma^0 + \frac{1}{\sqrt{6}} \Lambda &
\Sigma^+ & p \\
\Sigma^- & - \frac{1}{\sqrt{2}} \Sigma^0 + \frac{1}{\sqrt{6}} \Lambda & n \\
\Xi^- & \Xi^0 & - \frac{2}{\sqrt{6}} \Lambda
\end{array}
\right) \, .
\end{equation}
Unlike Eq.~\eqref{vpplag}, the $SU(4)$ extension of Eq.~\eqref{vbblag} is not trivial. However, 
in this work we follow the procedure adopted in Refs.~\cite{ourpaper,omegab}, which allows us 
to obtain in an easy manner the $VBB$ vertex without making use of $SU(4)$. In order to do this, 
we write the vector meson exchanged and the baryon in terms of quarks. For instance, let us 
consider that we have for the $VBB$ vertex a $\rho$ meson and two protons, that is the $\rho\,p\,p$ vertex. The $\rho^0$ meson wave function in terms of quarks is given by
\begin{equation}\label{rho}
\rho^0=\frac{1}{\sqrt{2}}(u\bar{u} - d\bar{d})\, .
\end{equation}
Inherent in the approach of Refs.~\cite{ourpaper,montana,romanets} is the neglect of the 
three momenta of the particles compared to the vector meson mass. This allows us to 
make the approximation $\gamma_{\mu}\to \gamma_0$, and hence applying Eq.~\eqref{rho} 
in the proton wave function as a number operator with a 
coupling $g$ provides us a spin independent operator at the quark level. In addition, we know 
that the proton wave function can be written as
\begin{equation}
p=\frac{1}{\sqrt{2}}|\phi_{MS}\chi_{MS} + \phi_{MA}\chi_{MA}\rangle \, ,
\end{equation}
where $\phi_{MS(MA)}$ and $\chi_{MS(MA)}$ are the flavor and spin mixed symmetric (antisymmetric) wave 
functions, respectively. Therefore, for the $\rho p p$ vertex we have
\begin{equation}\label{mxele}
\langle p | g \rho | p \rangle = \frac{g}{2} \langle \phi_{MS}\chi_{MS} + \phi_{MA}\chi_{MA}| \,
\frac{1}{\sqrt{2}}(u\bar{u} - d\bar{d})\, | \phi_{MS}\chi_{MS} + \phi_{MA}\chi_{MA} \rangle \, , 
\end{equation}
which provides us the same result as if we had used Eq.~\eqref{vbblag}. We use this method 
to evaluate the coupling of the vectors with baryons in the charm sector.

%Next, we present a list of all baryon wave functions used to obtain the transitions between the channels listed in Tables~\ref{tab1}, \ref{tab1new}, \ref{tab2} and \ref{tab2new}.

\subsection{Baryon wave functions}
\label{bwf}

In this subsection, we are going to write the baryon wave functions, which 
will be useful to evaluate the $VBB$ vertex. In order to 
obtain this vertex through the method we have discussed previously, where 
we have used the $\rho p p$ vertex as an example, we have to write down the spin-flavor 
wave functions for the baryons with $J^P=1/2^+$ and $J^P=3/2^+$. 
The spin-flavor wave functions associated with all of them are displayed in 
Tables~\ref{wave1over2} and \ref{wave3over2}. In those tables, 
the flavor part of the wave function is explicitly written leaving the heavy quarks 
as spectators and using $SU(3)$ symmetry in the light quarks, where the $SU(3)$ 
content can be mixed symmetric or mixed antisymmetric, $\phi_{MS(MA)}$, while 
the spin part, can be mixed symmetric, antisymmetric $\chi_{MS(MA)}$, or 
fully symmetric, $\chi_{S}$, as defined in the last column. 
\begin{table}[h!]
\caption{Wave functions for baryon with $J^P=1/2^+$.}
\centering
\begin{tabular}{c | c c c }
\hline\hline
{\bf States} ~& ~$I,\,J$~ & ~\textrm{Flavor}~ & ~\textrm{Spin}~($S_z=1/2$)\\
\hline
$\Lambda_c$~& ~$0,\,1/2$~&~ $\frac{c}{\sqrt{2}}(ud-du)$~ & ~$\frac{\uparrow}{\sqrt{2}}(\uparrow\downarrow -\downarrow\uparrow)$ \,\, $\chi_{MA}$\footnote{Mixed antisymmetric}\\
$\Sigma^+_c$~& ~$1,\,1/2$~& ~$\frac{c}{\sqrt{2}}(ud+du)$~ & ~$\frac{1}{\sqrt{6}}(\uparrow\downarrow\uparrow+\downarrow\uparrow\uparrow-2\uparrow\uparrow\downarrow)$\,\, $\chi_{MS}$ \footnote{Mixed symmetric}\\
$\Xi^+_{c}$~& ~$1/2,\,1/2$~& ~$\frac{c}{\sqrt{2}}(us-su)$~& ~$\frac{\uparrow}{\sqrt{2}}(\uparrow\downarrow-\downarrow\uparrow)$\,\, $\chi_{MA}$\\
$\Xi_c^{\prime\,+}$~&~$1/2,\,1/2$~&~$\frac{c}{\sqrt{2}}(us+su)$~&~$\chi_{MS}$\\
$\Xi^{++}_{cc}$~&~$1/2,\,1/2$~&~$ccu$~&~$\chi_{MS}$\\
$\Omega_{cc}$~&~$0,1/2$~&~$ccs$~&~$\chi_{MS}$\\
\hline\hline
\end{tabular}
\label{wave1over2}
\end{table}

\begin{table}[h!]
\caption{Wave functions for baryon with $J^P=3/2^+$.}
\centering
\begin{tabular}{c | c c c }
\hline\hline
{\bf States} ~& ~$I,\,J$~ & ~\textrm{Flavor}~ & ~\textrm{Spin}~($S_z=3/2$)\\
\hline
$\Sigma^{*\,+}_c$~&~$1,\,3/2$~&~$\frac{c}{\sqrt{2}}(ud+du)$~&~$\uparrow\uparrow\uparrow$\,\,$\chi_S$\footnote{Fully symmetric}\\
$\Xi^{*\,+}_c$~&~$1/2,\,3/2$~&~$\frac{c}{\sqrt{2}}(us+su)$~&~$\chi_S$\\
$\Omega^*_{cc}$~&~$1/2,\,3/2$~&~$ccs$~&~$\chi_S$\\
\hline\hline
\end{tabular}
\label{wave3over2}
\end{table}

Once we know the spin-flavor wave functions for the baryons, we can obtain the $VBB$ 
vertex, which corresponds to the lower one in Fig.~\ref{MBdiag}. For instance, 
if a $\Xi_{cc}$ is involved in a given transition, using its corresponding wave function defined in 
Table~\ref{wave1over2}, we can obtain the $\Xi^{++}_{cc}\Xi^{++}_{cc}\rho^0$ vertex evaluating the 
$\langle \Xi^{++}_{cc}  | g\,\rho^0 | \Xi^{++}_{cc} \rangle$ matrix element, which is
\begin{eqnarray}
\langle \Xi^{++}_{cc}  | g\,\rho^0 | \Xi^{++}_{cc} \rangle&=&g\langle ccu  | \otimes \langle \chi_{MS} |\Big[ \frac{1}{\sqrt{2}}
(u\bar{u}-d\bar{d})\Big]| \chi_{MS} \rangle \otimes | ccu \rangle\nonumber\\
&=&\frac{g}{\sqrt{2}}\, .
\end{eqnarray}
Analogously, for all the remaining $VBB$ vertices we are concerned, we will follow this 
procedure.

In order to match the $\Xi_{cc}$ isospin, we should construct states 
with $I=1/2$. We have the following multiplets
\begin{eqnarray}\label{multiplets}
\Xi_{cc}=\myvec{\Xi^{++}_{cc}\\ \Xi^+_{cc}};\,\, \Xi_c=\myvec{\Xi^+_c\\ \Xi^0_c};\,\, \Xi^{\prime}_c=\myvec{\Xi^{\prime\,+}_c\\ \Xi^{\prime\,0}_c};\,\, \Sigma_c=\myvec{-\Sigma_c^{++}\\ \Sigma^+_{c}\\\Sigma_c^0};\nonumber\\
D=\myvec{D^+\\-D^0};\,\, K=\myvec{K^+\\K^0};\,\, \pi=\myvec{-\pi^+\\ 	\pi^0\\ \pi^-},
\end{eqnarray}
from which we can obtain the $I=1/2$ states.

Now, we have all the elements needed to evaluate the transition amplitudes, 
depicted generically in Fig.~\ref{MBdiag}, 
between the relevant channels listed in Tables~\ref{tab1}, \ref{tab1new}, \ref{tab2} and \ref{tab2new} 
for each $J^P$ case. For $PB\to PB$ transitions by means of vector 
meson exchange, two vertices must be calculated. They are the $VPP$, that can be 
obtained from the Lagrangian defined in Eq.~\eqref{vpplag}, and the $VBB$, 
which is obtained employing the method we have just discussed. On the other hand, for 
$VB\to VB$ transitions the only difference is that now we have the $VVV$ 
vertex instead of $VPP$ one, and it can be evaluated using the following Lagrangian 
\begin{equation}\label{3vlag}
\mathcal{L}_{VVV}=ig\langle\,\, [V^{\mu},\partial_{\nu}V_{\mu}]\,V^{\nu}\,\rangle\,\, ,
\end{equation}
with the coupling $g$ the same as in Eq.~\eqref{vpplag}. In the case where 
the three-momentum of the vector meson is neglected versus the vector meson 
mass, as we also do here, only $\nu=0$ contributes in Eq.~\eqref{3vlag} 
which forces $V^{\nu}$ to be the exchanged vector, and the structure of the 
vertex is identical to the one of Eq.~\eqref{vpplag} for pseudoscalars, with the 
additional factor $\vec{\epsilon} \cdot \vec{\epsilon}^{\,\,\prime}$, with $\vec{\epsilon}$, 
$\vec{\epsilon}^{\,\,\prime}$ the polarization vectors of the external vector 
mesons \cite{hgmb1}. Following this procedure we can 
calculate all the transition amplitudes (see Appendix \ref{app} for more details), 
which have the same structure for every transition we are considering in this work, which is
\begin{equation}\label{vnr}
V_{ij}=C_{ij}\frac{1}{4f_{\pi}^2}(p^0 + p^{\prime \,0})\, ,
\end{equation}
where $p^0$ and $p^{\prime \,0}$ are the energies of the incoming and 
outgoing mesons, while the $C_{ij}$ are the coefficients, given in Table~\ref{tabCij}, 
for the pseudoscalar meson-baryon case with $J^P={1/2}^-$. Furthermore, 
the indices $i,\,j$ stand for the initial and final channels, respectively. For the vector 
meson-baryon, the coefficients are 
given in Table~\ref{tabDij}. The other cases are tabulated in 
Tables~\ref{tabEij} and \ref{tabFij}, respectively. Alternatively, we can also use the 
expression below
\begin{equation}\label{vrel}
V_{ij}=C_{ij}\frac{2\sqrt{s}-M_{B_{i}}-M_{B_{j}}}{4f_{\pi}^2}\sqrt{\frac{M_{B_{i}}+E_{B_{i}}}{2M_{B_{i}}}}
\sqrt{\frac{M_{B_{j}}+E_{B_{j}}}{2M_{B_{j}}}}\, ,
\end{equation}
which is obtained, according to Ref.~\cite{vijrel}, when we take into account 
relativistic corrections in $S$-wave. 

\begin{table}[h!]
\caption{$C_{ij}$ coeficients of Eq.~\eqref{vnr} for the pseudoscalar meson-baryon states coupling to $J^P={1/2}^-$ in $S$-wave.}
\centering
\begin{tabular}{c || c c c c c c c}
\hline\hline
 $PB_{1/2}$~ & ~~$\Xi_{cc}\pi$~ & ~$\Lambda_c D$~ & ~$\Xi_{cc} \eta$~ & ~$\Omega_{cc} K$ ~&~ $\Sigma_c D$ ~& ~$\Xi_c D_s$ ~& ~$\Xi^{\prime}_cD_s$\\
\hline\hline
$\Xi_{cc}\pi$ & $-\frac{4}{3}$ & $0$ & $-\frac{\sqrt{2}}{3}$ & $-\sqrt{\frac{3}{2}}$ & $0$ & $0$ & $0$\\
$\Lambda_c D$ &  & $-1$ & $0$ & $0$ & $0$ & $-1$ & $0$\\
$\Xi_{cc} \eta$ & & & 0 & $-\frac{1}{\sqrt{3}}$ & $0$ & $0$ & $0$\\
$\Omega_{cc} K$ & & &  & $-1$ & $0$ & $0$ & $0$\\
$\Sigma_c D$ & & & & & $-3$ & $0$ & $-\frac{1}{\sqrt{3}}$\\
$\Xi_c D_s$ & & & & & & $-1$ & $0$\\
$\Xi^{\prime}_cD_s$ & & & & & & & $-1$\\
\hline\hline
\end{tabular}
\label{tabCij}
\end{table}

\begin{table}[h!]
\caption{$C_{ij}$ coeficients of Eq.~\eqref{vnr} for the vector meson-baryon states coupling to $J^P={1/2}^-,\, 3/2^-$ in $S$-wave.}
\centering
\begin{tabular}{c || c c c c c c c c}
\hline\hline
 $VB_{1/2}$~ & ~~$\Lambda_{c}D^*$~ & ~$\Xi_{cc} \rho$~ & ~$\Xi_{cc} \omega$~ & ~$\Sigma_{c} D^*$ ~& ~$\Xi_c D_s^*$ ~& ~$\Omega_{cc} K^*$ ~&~ $\Xi_{cc} \phi$ ~&~ $\Xi_c^{\prime} D_s^*$\\
\hline\hline
$\Lambda_{c}D^*$ & $-1$ & $0$ & $0$ & $0$ & $-1$ & $0$ & $0$ & $0$\\
$\Xi_{cc} \rho$ &  & $-\frac{4}{3}$ & $-\frac{1}{\sqrt{3}}$ & $0$ & $0$ & $-\sqrt{\frac{3}{2}}$ & $0$ & $0$\\
$\Xi_{cc} \omega$ & & & $0$ & $0$ & $0$ & $-\frac{1}{\sqrt{2}}$ & $0$ & $0$\\
$\Sigma_{c} D^*$ & & & & $-3$ & $0$ & $0$ & $0$ & $-\frac{1}{\sqrt{3}}$\\
$\Xi_c D^*_s$ & & & & & $-1$ & $0$ & $0$ & $0$\\
$\Omega_{cc} K^*$ & & & & & & $-1$ & $1$ & $0$\\
$\Xi_{cc} \phi$ & & & & & & & $0$ & $0$\\
$\Xi^{\prime}_c D^*_s$ & & & & & & & & $-1$\\
\hline\hline
\end{tabular}
\label{tabDij}
\end{table}

\begin{table}[h!]
\caption{$C_{ij}$ coeficients of Eq.~\eqref{vnr} for the pseudoscalar meson-baryon states coupling to $J^P={3/2}^-$ in $S$-wave.}
\centering
\begin{tabular}{c || c c c c c}
\hline\hline
 $PB_{3/2}$~ & ~~$\Xi^*_{cc}\pi$~ & ~$\Xi^*_{cc} \eta$~ & ~$\Omega^*_{cc} K$~ & ~$\Sigma^*_{c} D$ ~&~ $\Xi^*_c D_s$\\
\hline\hline
$\Xi^*_{cc}\pi$ & $-\frac{4}{3}$ & $-\frac{\sqrt{2}}{3}$ & $-\sqrt{\frac{3}{2}}$ & $0$ & $0$\\
$\Xi^*_{cc} \eta$ & & $0$ & $-\frac{1}{\sqrt{3}}$ & $0$ & $0$\\
$\Omega^*_{cc} K$ & & & $-1$ & $0$ & $0$\\
$\Sigma^*_{c} D$ & & & & $-3$ & $-\frac{1}{\sqrt{3}}$\\
$\Xi^*_c D_s$ & & & & & $-1$\\
\hline\hline
\end{tabular}
\label{tabEij}
\end{table}

\begin{table}[h!]
\caption{$C_{ij}$ coeficients of Eq.~\eqref{vnr} for the vector meson-baryon states coupling to $J^P={1/2}^-,\,{3/2}^-,\, {5/2}^-$ in $S$-wave.}
\centering
\begin{tabular}{c || c c c c c c}
\hline\hline
 $VB_{3/2}$~ & ~~$\Xi^*_{cc}\rho$~ & ~$\Xi^*_{cc} \omega$~ & ~$\Sigma^*_{c} D^*$~ & ~$\Omega^*_{cc} K^*$ ~&~ $\Xi^*_{cc} \phi$ ~&~ $\Xi^*_{c} D^*_s$\\
\hline\hline
$\Xi^*_{cc}\rho$ & $-\frac{4}{3}$ & $-\frac{1}{\sqrt{3}}$ & $0$ & $-\sqrt{\frac{3}{2}}$ & $0$ & $0$\\
$\Xi^*_{cc} \omega$ & & $0$ & $0$ & $-\frac{1}{\sqrt{2}}$ & $0$ & $0$\\
$\Sigma^*_{c} D^*$ & & & $-3$ & $0$ & $0$ & $-\frac{1}{\sqrt{3}}$\\
$\Omega^*_{cc} K^*$ & & & & $-1$ & $1$ & $0$\\
$\Xi^*_{cc} \phi$ & & & & & $0$ & $0$\\
$\Xi_{c}^* D_s^*$ & & & & & & $-1$\\
\hline\hline
\end{tabular}
\label{tabFij}
\end{table}

%%%%%%%%%%%%%%%%%%%%%%%%%%%%%%%%%%%%%%%%%%%%%%%%%%%%%%%%%%%%%%%%%%%%%%%%%%%%%%%%%%% 
\subsection{The scattering matrix for meson-baryon interaction}
%%%%%%%%%%%%%%%%%%%%%%%%%%%%%%%%%%%%%%%%%%%%%%%%%%%%%%%%%%%%%%%%%%%%%%%%%%%%%%%%%%%

In order to unitarize the transition amplitudes, often called potentials, given by 
Eqs.~\eqref{vnr} or \eqref{vrel}, we have to use them as the kernel of the 
Bethe-Salpeter equation, where poles in resulting scattering matrices correspond to bound 
states or resonances. The Bethe-Salpeter equation in its on-shell factorization 
form \cite{chuapp1,chuapp2,lamb3} is defined as 
\begin{equation}\label{bs}
T = (1-VG)^{-1}\,V\, ,
\end{equation}
where $V$ is the matrix that describes the transition amplitudes between each 
channel listed in Tables \ref{tab1}, \ref{tab1new}, \ref{tab2} and \ref{tab2new}. 
$G$ is the meson-baryon loop 
function, which is divergent. It is possible to evaluate this function by means 
of dimensional regularization or with a three momentum cutoff. In this work, 
we use the cutoff method since the dimensional regularization in the 
heavy sector might lead to solutions that are not related to 
physical states \cite{gpositive}. This happens because in that scheme of regularization the loop 
function can assume positive values below threshold. As a result, poles might 
manifest in the solution even with the potential being repulsive.
In the cutoff method the meson-baryon loop 
function is given by
 \begin{eqnarray}\label{loop}
 G_l &=& i \int \frac{d^4q}{(2\pi)^4} \frac{M_l}{E_l({\bf q})}\frac{1}{k^0+p^0 - q^0 - E_l({\bf q})+i\epsilon} \frac{1}{{\bf q}^2-m^2_l+i\epsilon}\nonumber\\
 &=& \int_{|{\bf q}|< q_{max}} \frac{d^3{\bf q}}{(2\pi)^3}\frac{1}{2\omega_l({\bf q})}\frac{M_l}{E_l({\bf q})}\frac{1}{k^0+p^0-\omega_l({\bf q}) - E_l({\bf q})+i\epsilon}\, ,
 \end{eqnarray}
where the subscript $l$ is the label for the $l$th-channel, while $k^0+p^0=\sqrt{s}$ and 
$\omega_l$, $E_l$ stand for the meson and baryon energies, respectively. In 
the next section, we show the results for $q_{max}=650$ MeV, which is the same 
value used in previous works \cite{ourpaper,omegab}, where the same framework discussed 
here was applied to investigate meson-baryon interactions in the heavy sector. It was 
shown in Refs.~\cite{hqss1,hqss2} that the same value of cutoff has to be used for all channels in order to respect the rules of heavy quark symmetry.

The bound states and resonances can be associated to the poles that are 
solutions of the Bethe-Salpeter equation, defined in Eq.~\eqref{bs}. In order 
to look for these poles we need to obtain the $T$-matrix in the complex energy plane, for which 
we have calculated the meson-baryon $G$ function in the first (\textrm{I}) and 
second (\textrm{II}) Riemann sheet \cite{chuapp1}. This is done by changing $G$ in 
Eq.~\eqref{bs} to $G^{II}$ in order to obtain $T^{II}$. The loop $G^{II}$ 
is the analytic continuation of the loop function in the second Riemann sheet, and 
it is given by
\begin{eqnarray}\label{gii}
G^{II}_l(\sqrt{s})&=&G^{I}_l(\sqrt{s}) + i\frac{M_l}{2\pi \sqrt{s}}\,p\, , \,\, \textrm{with} \,\, Im(p)>0,\nonumber\\
p&=&\frac{\lambda^{1/2}(s,m^2_l,M^2_l)}{2\sqrt{s}}\, ,
\end{eqnarray}
with $m_l$ and $M_l$ being the meson and baryon masses of the $l$-channel, respectively, while $G^{I}$ (given by Eq.~\eqref{loop}) 
and $G_l^{II}$ stand for the loop function in the first and second Riemann 
sheet, respectively. In Eq.~\eqref{gii}, we use $G_l^{II}$ when the $l$th-channel 
is open, i. e. $Re(\sqrt{s}\,)> m_l + M_l$. On the other hand, when the channel 
is closed, that is $Re(\sqrt{s}\,) < m_l + M_l$, we have $G_l^{II}=G_l^{I}$.

It is also possible to evaluate the couplings $g_l$ of the state to the different 
meson-baryon channels. In order to do this, note that close to the pole 
the amplitude in the complex plane for a diagonal transition can be written as
\begin{equation}\label{gl}
T_{ll}(s)\approx \frac{g_l^2}{\sqrt{s}-z_R}\, ,
\end{equation}
where $z_R=M_R+i\Gamma_R/2$ stands for the position of the bound state/resonance 
\cite{javi}. Hence, the 
coupling can be evaluated as the residue at the pole of $T_{ll}(s)$, by 
means of the following formula
\begin{equation}
g_l^2=\frac{r}{2\pi}\int\limits_0^{2\pi}\, T_{ll}(z(\theta)) e^{i\theta} d\theta \, ,
\end{equation}
where $z=z_R+re^{i\theta}$.

In addition, with the coupling constant and the $G$ function calculated at the 
pole, we can obtain $g_l\,G_l(z_R)$, which is proportional to the wave function at the origin in the $l$th-channel \cite{couplings}.

%%%%%%%%%%%%%%%%%%%%%%%%%%%%%%%%%%%%%%%%%%%%%%%%%%%%%%%%%%%%%%%%%%%%%%%%%%%%%%%%%%% 
\section{Results}
%%%%%%%%%%%%%%%%%%%%%%%%%%%%%%%%%%%%%%%%%%%%%%%%%%%%%%%%%%%%%%%%%%%%%%%%%%%%%%%%%%%

In Table~\ref{tab:gB1/2P} we show the poles we have found according to the 
procedure discussed previously. They are related to the interaction involving a 
pseudoscalar meson and $1/2^+$ baryon in $S$-wave, such that for this case we have poles associated to the $J^P=1/2^-$ quantum numbers. In addition, we also show the couplings of these 
states to the channels spanning the space of states listed in Table~\ref{tab1} 
as well as the product $g_lG^{II}_l$, with  $G^{II}_l$ being the loop function 
evaluated at the pole in the second Riemann sheet. We get two states separated approximately by 
$\approx10$ MeV, with one at $4082.79$ MeV and the other at $4092.20$ MeV. From 
the results obtained for the couplings as well as for the wave function at the origin, we 
observe that the first pole couples strongly to the $\Sigma_cD$ 
channel. It also couples to $\Xi_c^{\prime}D_s$, but with a smaller value than to the 
former channel. We can understand this by looking at Table~\ref{tabCij}. According to that table 
only the $\Sigma_cD\to \Sigma_cD$ and $\Sigma_cD\to \Xi^{\prime}_cD_s$ transitions 
are allowed, with the coefficient related to the diagonal one as the biggest value. Therefore, 
we can say that this pole is mostly a $\Sigma_cD$ molecule.
\begin{table}[h!]
\caption{Poles and couplings in the $PB_{1/2}$, $J^P={1/2}^-$ sector, with $q_{max}=650$ MeV, and $g_l\,G^{II}_l$ in MeV.}
\centering
\begin{tabular}{c c c c c c c c c}
\hline\hline
${\bf 4082.79}$ & ~$\Xi_{cc}\pi$~ &  ~$\Lambda_c D$~ & ~$\Xi_{cc}\eta$~ & ~$\Omega_{cc} K$ ~&~ $\Sigma_{c} D$ ~& ~$\Xi_c D_s$ ~& ~$\Xi^{\prime}_c D_s$\\
\hline
$g_l$ & $0$ & $0$ & $0$ & $0$ & $\bf 8.86$ & $0$ & $1.93$ & \\ %Main channel $=5$ (real $g_5>0$)
$g_l\,G^{II}_l$ & $0$ & $0$ & $0$ & $0$ & $\bf -31.29$ & $0$ & $-4.04$ & \\ %Main channel $=2$ (real $g_2>0$)
\hline\hline
${\bf 4092.20}$ & ~$\Xi_{cc}\pi$~ &  ~$\Lambda_c D$~ & ~$\Xi_{cc}\eta$~ & ~$\Omega_{cc} K$ ~&~ $\Sigma_{c} D$ ~& ~$\Xi_c D_s$ ~& ~$\Xi^{\prime}_c D_s$\\
\hline
$g_l$ & $0$ & $\bf 4.01$ &  $0$ & $0$ & $0$ & $3.75$ & $0$ \\
$g_l\,G^{II}_l$ & $0$ & $\bf -29.49$ & $0$ & $0$ & $0$ & $-9.76$ &  $0$ \\
%\hline\hline
%${\bf\color{red} 4518.94 +i0.37}$ & ~$\Xi_{cc}\pi$~ &  ~$\Lambda_c D$~ & ~$\Xi_{cc}\eta$~ & ~$\Omega_{cc} K$ ~&~ $\color{red}\Sigma_{c} D$ ~& ~$\Xi_c D_s$ ~& ~$\Xi^{\prime}_c D_s$\\
%\hline
%$g_i$ & $0$ & $0$ &  $0$ & $0$ & $0.04 -i0.07$ & $0$ & $\bf 2.89$ \\
%$g_i\,G^{II}_i$ & $0$ & $0$ & $0$ & $0$ & $5.23 +i0.22$ & $0$ &  $\bf -28.94 -i0.14$ \\ %Main channel $=7$ (real $g_7>0$)
\hline\hline
\end{tabular}
\label{tab:gB1/2P}
\end{table}
For the second pole, at $4092.20$ MeV, we see that it couples to both $\Lambda_cD$ 
and $\Xi_cD_s$ channels with almost the same magnitude. The only open channel for 
both states found is $\Xi_{cc}\pi$, but as can be seen from Table~\ref{tab:gB1/2P}, they 
do not couple to this channel.
 
\begin{table}[h!]
\caption{Poles and couplings in the $PB_{3/2}$, $J^P={3/2}^-$ sector, with $q_{max}=650$ MeV, and $g_l\,G^{II}_l$ in MeV.}
\centering
\begin{tabular}{c c c c c c}
\hline\hline
${\bf 4149.67}$ &~$\Xi^*_{cc}\pi$~ & ~$\Xi^*_{cc}\eta$~ & ~$\Omega^*_{cc} K$ ~&~ $\Sigma^*_{c} D$ ~& ~$\Xi^*_c D_s$~ \\
\hline
$g_l$ & $0$ & $0$ & $0$ & $\bf8.82$ & $1.30$ \\
$g_l\,G^{II}_l$ & $0$ & $0$ & $0$ & $\bf-31.46$ & $-2.71$ \\
%\hline\hline
%${\bf \color{red} 4574.94 +i0.22}$ &~$\Xi^*_{cc}\pi$~ & ~$\Xi^*_{cc}\eta$~ & ~$\Omega^*_{cc} K$ ~&~ $\color{red}\Sigma^*_{c} D$ ~& ~$\Xi^*_c D_s$~ \\
%\hline
%$g_i$ & $0$ & $0$ & $0$ & $0.07 -i0.06$ & $\bf 3.42 -i0.05$ \\
%$g_i\,G^{II}_i$ & $0$ & $0$ & $0$ & $3.8 +i0.10$ & $\bf -30.08 +i0.39$ \\
\hline\hline
\end{tabular}
\label{tab:gB3/2P}
\end{table}

The results associated with the interaction 
involving a pseudoscalar meson and a baryon with $J^P=3/2^+$ in $S$-wave, are displayed in Table~\ref{tab:gB3/2P}. Analogously to the previous case, we also present the couplings together with the wave function at the origin, that is the $g_l G^{II}_l$ product. In this case, we have just one 
pole at $4149.67$ MeV, coupling mostly to $\Sigma^*_cD$ and with less intensity to the $\Xi^*_cD_s$ channel. As can be seen looking at the Table~\ref{tab2} for the thresholds, only the $\Xi^*_{cc}\pi$ channel is open. The coupling to $\Sigma^*_cD$ is almost seven times bigger than the value for the 
other channel, $\Xi^*_cD_s$, then this pole is naturally associated with a $\Sigma^*_cD$ molecule. 
This pole would be the spin partner of the pole found at $4082.79$ MeV from the 
pseudoscalar-baryon ($PB$) interaction, with $J^P=1/2^-$.

Next we look for the states with degenerate $J^P=1/2^-,\, 3/2^-$, resulting from the interaction in 
$S$-wave of vector mesons and baryons with $J^P=1/2^+$. Our findings for this particular case can be seen in Table~\ref{tab:gB1/2V}. 
Three states have been found, at $4217.21$ MeV, $4229.19$ MeV and at $4328.65$ MeV. 
The first of them couples strongly to $\Sigma_c D^*$ and little to the $\Xi^{\prime}_c D^*_s$ channel, 
and hence, this pole qualifies as a $\Sigma_c D^*$ bound state. The second state found, 
at $4229.19$ MeV, couples to both $\Lambda_cD^*$ and $\Xi_c D^*_s$ with similar values 
for the couplings, however, when we compare the values for the product $g_l G_l^{II}$, we 
see that the one for the $\Lambda_cD^*$ channel is much bigger than that for $\Xi_c D^*_s$. The same behavior is found for the last pole, at $4328.65$ MeV, whose couplings to $\Xi_{cc} \rho$ and 
to $\Omega_{cc}K^*$ are of the same order, while the value for the wave function at the origin 
for the former channel is bigger than that for the latter one. It is worth mentioning that three states 
were also obtained in the same $VB_{1/2}$, $J^P=1/2^-, 3/2^-$ sector for the $\Omega_c$ and 
$\Omega_b$ states, respectively, studied in Refs.~\cite{ourpaper,omegab}. We note that for the 
first and second poles at $4217.21$ MeV and $4229.19$ MeV, all channels are closed for decay. 
The third pole at $4328.65$ MeV has about $30$ MeV of phase space to decay into 
$\Lambda_cD^*$, however in our approach it does not couple to this channel since it would 
require the exchange of a heavy vector meson.

\begin{table}[h!]
\caption{Poles and couplings in the $VB_{1/2}$, $J^P={1/2}^-, \, {3/2}^- $ sector, with $q_{max}=650$ MeV, and $g_l\,G^{II}_l$ in MeV.}
\centering
\begin{tabular}{c c c c c c c c c}
\hline\hline
${\bf 4217.21}$ & ~$\Lambda_c D^*$~ & ~$\Xi_{cc}\rho$~ & ~$\Xi_{cc}\omega$~ & ~$\Sigma_{c} D^*$  ~& ~$\Xi_c D^*_s$ ~ & ~ $\Omega_{cc} K^*$~&~ $\Xi_{cc}\phi$ ~ & ~ $\Xi^{\prime}_c D^*_s$ ~\\
\hline
$g_l$ &  $0$ & $0$ & $0$ & $\bf 9.31$ & $0$ & $0$ & $0$ & $2.03$ \\
$g_l\,G^{II}_l$ &  $0$ & $0$ & $0$ & $\bf -30.40$ & $0$ & $0$ & $0$ & $-3.94$ \\
\hline\hline
${\bf 4229.19}$ & ~$\Lambda_c D^*$~ & ~$\Xi_{cc}\rho$~ & ~$\Xi_{cc}\omega$~ & ~$\Sigma_{c} D^*$  ~& ~$\Xi_c D^*_s$ ~ & ~ $\Omega_{cc} K^*$~&~ $\Xi_{cc}\phi$ ~ & ~ $\Xi^{\prime}_c D^*_s$ ~\\
\hline
$g_l$ &  $\bf 4.21$ & $0$ & $0$ & $0$ & $3.98$ & $0$ & $0$ & $0$ \\
$g_l\,G^{II}_l$ &  $\bf -28.70$ & $0$ & $0$ & $0$ & $-9.59$ & $0$ & $0$ & $0$ \\
\hline\hline
${\bf 4328.65}$ & ~$\Lambda_c D^*$~ & ~$\Xi_{cc}\rho$~ & ~$\Xi_{cc}\omega$~ & ~$\Sigma_{c} D^*$  ~& ~$\Xi_c D^*_s$ ~ & ~ $\Omega_{cc} K^*$ ~ &~ $\Xi_{cc}\phi$ ~&~ $\Xi^{\prime}_c D^*_s$ ~\\
\hline
$g_l$ &  $0$ & $\bf 2.95$ & $1.23$ & $0$ & $0$ & $2.66$ & $-0.56$ & $0$ \\
$g_l\,G^{II}_l$ &  $0$ & $\bf -35.46$ & $-14.21$ & $0$ & $0$ & $-14.72$ & $2.64$ & $0$ \\
%\hline\hline
%${\bf \color{red}  4660.31 +i0.42}$ & ~$\Lambda_c D^*$~ & ~$\Xi_{cc}\rho$~ & ~$\Xi_{cc}\omega$~ & ~$\color{red}\Sigma_{c} D^*$ ~&~ $\Xi_{cc}\phi$ ~& ~$\Xi_c D^*_s$ ~ & ~ $\Omega_{cc} K^*$ ~ & ~ $\Xi^{\prime}_c D^*_s$ ~\\
%\hline
%$g_i$ &  $0$ & $0$ & $0$ & $0.03 - i0.08$ & $0$ & $0$ & $0$ & $\bf 3.03 -i0.01$ \\
%$g_i\,G^{II}_i$ &  $0$ & $0$ & $0$ & $5.16 +i0.21$ & $0$ & $0$ & $0$ & $\bf-28.30$ \\
\hline\hline
\end{tabular}
\label{tab:gB1/2V}
\end{table}

\begin{table}[h!]
\caption{Poles and couplings in the $VB_{3/2}$, $J^P={1/2}^-, \, {3/2}^-, \, {5/2}^- $ sector, with $q_{max}=650$ MeV, and $g_l\,G^{II}_l$ in MeV.}
\centering
\begin{tabular}{c c c c c c c}
\hline\hline
${\bf 4280.43}$ & ~$\Xi^*_{cc}\rho$~ & ~$\Xi^*_{cc}\omega$~ & ~$\Sigma^*_{c} D^*$ ~&  $\Omega^*_{cc} K^*$ & ~$\Xi^*_{cc}\phi$ ~ & ~$\Xi^*_c D^*_s$ ~ \\
\hline
$g_l$ & $0$ & $0$ & $\bf 9.31$ & $0$ & $0$ & $2.03$ \\
$g_l\,G^{II}_l$ & $0$ & $0$ & $\bf -30.42$ & $0$ & $0$ & $-3.90$ \\
\hline\hline
${\bf 4409.61}$ & ~$\Xi^*_{cc}\rho$~ & ~$\Xi^*_{cc}\omega$~ & ~$\Sigma^*_{c} D^*$ ~& $\Omega^*_{cc} K^*$ & ~$\Xi^*_{cc}\phi$ ~ & ~$\Xi^*_c D^*_s$ ~ \\
\hline
$g_l$ & $\bf 2.95$ & $1.23$ & $0$ & $2.65$ & $-0.56$ & $0$ \\
$g_l\,G^{II}_l$ & $\bf -35.51$ & $-14.22$ & $0$ & $-14.63$ & $2.62$ & $0$ \\
\hline\hline
\end{tabular}
\label{tab:gB3/2V}
\end{table}

Finally we also show in Table~\ref{tab:gB3/2V} the results for the vector 
meson-baryon states, with $J^P=3/2^+$ for the baryon. In this case, 
we obtain two poles: $4280.43$ MeV and $4409.61$ MeV. The first one 
couples strongly to $\Sigma_c^*D^*$ and since its coupling to the other 
channel, $\Xi_c^*D^*_s$, is four times smaller than the first one, this 
pole is likely a $\Sigma_c^*D^*$ molecule. On the other hand, the 
second pole couples almost to all channels, except for the 
$\Sigma_c^*D^*$ and $\Xi_c^*D^*_s$. It couples with similar values 
for the coupling to the channels: $\Xi_{cc}^*\rho$ and $\Omega_{cc}^*K^*$; 
and next to $\Xi_{cc}^*\omega$, and a little to $\Xi_{cc}^*\phi$. But, 
by looking at the wave function at the origin we conclude that this last 
pole comes mostly from the $\Xi_{cc}^*\rho$ channel.

Evidence of three resonances at higher energies has also been found. In the 
$B_{1/2}P$ sector a state coupling mostly to $\Xi_c^\prime D_s$ was found 
around 4520 MeV. This state also couples to $\Sigma_c D$, and would be the
 ``heavy partner'' of the pole found around 4080 MeV. However, the pole is 
 close to the threshold of $\Xi_c^\prime D_s$, which is about 200 MeV above 
 the one of $\Sigma_c D$. At this energy the propagator of $\Sigma_c D$ is 
 already too far from its threshold and its real part becomes positive, what 
 can affect the unitarization of the amplitude and yield unreliable results. The 
 same happens in the $B_{1/2}P$ sector, where a state coupling mostly to 
 $\Xi_c^\ast D_s$, and also to $\Sigma_c^\ast D$, was found around 4575 
 MeV; and in the $B_{1/2}V$ sector, where a state coupling mostly to 
 $\Xi_c^\prime D_s^\ast$, and also to $\Sigma_c D^\ast$, was found 
 around 4660 MeV. In order to be sure these poles have physical meaning 
 and do not come from the influence of the lower channel, we have repeated the 
 calculation using only the single dominant channel ($\Xi^{\prime}_cD_s$, 
 $\Xi^*_c D_s$ and $\Xi^{\prime}_cD^*_s$, respectively) and we have found 
 that the resonances are still present moving $10$ MeV or less from the previous 
 pole position. Therefore, these poles have indeed physical meaning; the only 
 difference is that their pole position has larger uncertainties in comparison with the 
 results presented in Tables~\ref{tab:gB1/2P}, \ref{tab:gB3/2P}, \ref{tab:gB1/2V} and \ref{tab:gB3/2V}.

%%%%%%%%%%%%%%%%%%%%%%%%%%%%%%%%%%%%%%%%%%%%%%%%%%%%%%%%%%%%%%%%%%%%%%%%%%%%%%%%%%% 
\section{Conclusions}
%%%%%%%%%%%%%%%%%%%%%%%%%%%%%%%%%%%%%%%%%%%%%%%%%%%%%%%%%%%%%%%%%%%%%%%%%%%%%%%%%%%

Using a framework employed to study the $\Omega_c$ states, recently observed 
by the LHCb collaboration, and also used to predict similar structures in the beauty sector, 
named as $\Omega_b$ states, which are dynamically generated through meson-baryon 
interaction, we have investigated the dynamical generation of possible doubly charmed 
heavy baryons resonances in the $C=2$, $S=0$ and $I=1/2$ sector. In particular, the 
transition amplitudes between the relevant channels are inspired in the Lagrangians 
from the hidden local gauge approach. In order to deal with the $VBB$ vertex, the 
spin-flavor wave functions for the baryons were constructed considering the heavy quarks 
as spectators, allowing us to extend the $VBB$ 
interaction to the heavy sector in an easy manner, using the $SU(3)$ content of $SU(4)$. 

These transition amplitudes are unitarized taking them as the kernel of the Bethe-Salpeter 
equation, whose solutions can be associated with physical states. Using this approach, we 
have just one parameter, which is the regulator in the loop function of the meson-baryon 
states. We have used the cutoff regularization method with the same cutoff used in previous works 
in which we have employed this method to study other meson-baryon interactions in the 
beauty sector as well as to study singly charmed baryon resonances, recently observed 
by the LHCb and called $\Omega_c$ states.

We obtain two poles, one at $4083$ MeV and at $4092$ MeV, for pseudoscalar 
meson-baryon interaction with $J^P=1/2^-$, and another one at $4150$ MeV for 
$J^P=3/2^-$ quantum numbers. Furthermore, we have also considered the degenerate 
cases, stemming from the interaction of vector meson-baryon, that provides three states 
with $J^P=1/2^-$ and $3/2^-$: at $4217$ MeV, $4229$ MeV and $4329$ MeV, 
and two more with $J^P=1/2^-$, $3/2^-$ and $5/2^-$, at $4280$ MeV and at $4410$ 
MeV. 
%We have also found evidence associated with two states from baryon($1/2$)-vector meson 
%interaction ($J^P=1/2^-,\,3/2^-$): 

Therefore, we have predicted eight states, which in our approach are dynamically 
generated from the meson-baryon interaction. The molecular picture provides important 
ingredients on the $J^P$ quantum numbers of these states together with their quark 
structure and the observation of such states can 
give support to the molecular nature of these resonances as well as should serve as a 
good test for the extrapolations used of the chiral unitary approach.

\section*{Acknowledgments}

J.~M.~Dias would like to thank the Brazilian funding agency FAPESP for the financial support 
under Grant No. $2016/22561{\rm-}2$.
V.~R.~Debastiani wishes to acknowledge the support from the
Programa Santiago Grisolia of Generalitat Valenciana (Exp. GRISOLIA/2015/005).
This work is also partly supported by the Spanish Ministerio de Economia
y Competitividad and European FEDER funds under the contract number
FIS2014-57026-REDT, FIS2014-51948-C2-1-P, and FIS2014-51948-C2-2-P, and
the Generalitat Valenciana in the program Prometeo II-2014/068. It is also partly 
supported by the National Natural Science Foundation of China (Grants No. $11475227$ 
and No. $11735003$) and the Youth Innovation Promotion Association CAS (No. $2016367$).

\appendix
\section{An example of evaluation of the transition matrix elements for the meson-baryon channels}
\label{app}

In what follows we shall illustrate how to obtain the transition amplitudes, represented 
by the diagram depicted in Fig.~\ref{MBdiag}, using the $\Xi_{cc}\pi\to \Xi_{cc}\pi$ 
amplitude as an example. In this case, we have to consider that, in Fig.~\ref{MBdiag}, 
the initial $M_i$ and final $M_f$ meson will be a pion, while $B_i$ and $B_f$ are 
the corresponding initial and final baryon, that in this example will be the $\Xi_{cc}$. 
The vector meson exchanged $V$ could be a $\rho$ and an $\omega$ meson. 
However, in this particular case, there is no $\omega$ meson contribution since 
this vector meson does not couple to two pions.
\begin{figure}[h!]
	\begin{center} 
		\includegraphics[width=1.0\textwidth]{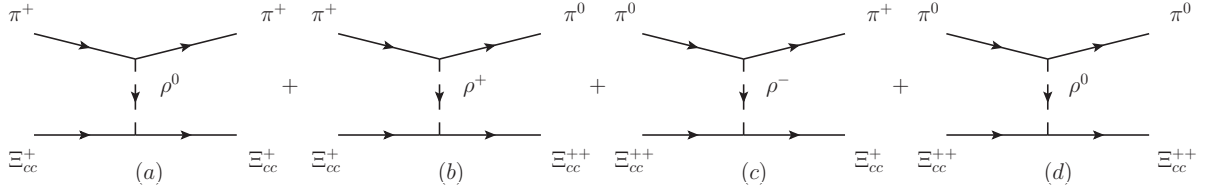}
	\end{center} 
	\caption{\label{mbint} Diagrams representing the meson-baryon interaction through vector 
	meson exchange for the particular $\Xi_{cc}\pi\to\Xi_{cc}\pi$ case. In this case, the vector meson 
	exchanged can be a $\rho^0$, $\rho^+$ or $\rho^-$.}
\end{figure}

Firstly, we have to write the $| \Xi_{cc}\pi \rangle$ in the $I=1/2$ combination. Remembering 
the multiplets defined in Eq.~\eqref{multiplets}, we have
\begin{equation}\label{iscomb}
|\, \Xi_{cc}\pi\, (I_3=1/2) \rangle = \sqrt{\frac{2}{3}}\, \Xi^+_{cc}\,\pi^+
 + \sqrt{\frac{1}{3}}\, \Xi^{++}_{cc}\,\pi^0 \,.
\end{equation}
Therefore, the transition $T_{\Xi_{cc}\pi\to\Xi_{cc}\pi}$ (before the unitarization of 
Eq.~\eqref{bs}) will be given evaluating the 
matrix element $T^{I=1/2}_{\Xi_{cc}\pi\to\Xi_{cc}\pi}=\langle\, \Xi_{cc}\pi\, (I_3=1/2) |\,T\,|\, \Xi_{cc}\pi\, (I_3=1/2)\, \rangle$, 
that results
\begin{equation}\label{amptotal}
T^{I=1/2}_{\Xi_{cc}\pi\to\Xi_{cc}\pi}=\frac{2}{3}\,t_{\Xi^{+}_{cc}\pi^+\to\Xi^{+}_{cc}\pi^+}
+2\sqrt{\frac{2}{3}}\,\sqrt{\frac{1}{3}}\,t_{\Xi^+_{cc}\pi^+\to \Xi^{++}_{cc}\pi^0}+
\frac{1}{3}\,t_{\Xi^{++}_{cc}\pi^0\to\Xi^{++}_{cc}\pi^0}\, .
\end{equation}
Thus, the task to get the transition we are interested in involves the calculation 
of each of the amplitudes appearing in Eq.~\eqref{amptotal}. The diagrams 
representing each of them are depicted in Fig.~\ref{mbint}. In order to calculate each 
diagram, we need to obtain all the corresponding upper and lower vertices.

The upper $VPP$ vertices in Fig.~\ref{mbint}, are obtained using the Lagrangian in 
Eq.~\eqref{vpplag}. We get
\begin{eqnarray}\label{upvert}
-it_{\pi^+\to \pi^+}\myvec{\rho^0\\ \omega}&=&2igV_{\mu}(p+p^{\prime})^{\mu}\myvec{1/\sqrt{2}\\0}\, ,\nonumber\\
-it_{\pi^+\to \pi^0\rho^+}&=& -g\frac{i}{\sqrt{2}} \rho^{+\mu}(p+p^{\prime})^{\mu}\, ,\nonumber\\
-it_{\pi^0\to \pi^0}\myvec{\rho^0\\ \omega}&=&0\, .
\end{eqnarray}

The next step focus on the calculation of the $VBB$ vertices. In this case, 
we shall follow the procedure described in Subsection~\ref{bwf}, in which 
we have shown how to get the 
amplitudes using the spin-flavor baryon wave functions. The lower vertices in 
Fig.~\ref{mbint} involve the $\Xi^+_{cc}$ and $\Xi^{++}_{cc}$ spin-flavor wave functions, and from 
Table~\ref{wave1over2} we have
\begin{eqnarray}\label{lowvert}
|\,\Xi^{++}_{cc}\,\rangle &=& | ccu \rangle \otimes | \chi_{MS} \rangle\\
|\,\Xi^+_{cc}\,\rangle &=& | ccd \rangle \otimes | \chi_{MS} \rangle\, .
\end{eqnarray}
Therefore, the vertices $\Xi^{+}_{cc}\Xi^{+}_{cc}\rho^0$ and 
$\Xi^{+}_{cc}\Xi^{++}_{cc}\rho^+$ are
\begin{equation}\label{lowvert1}
\langle\, ccd \,| \frac{g}{\sqrt{2}}(u\bar{u}-d\bar{d})|\, ccd \,\rangle 
\langle \chi_{MS}| \chi_{MS}\rangle=-\frac{g}{\sqrt{2}}\, ,
\end{equation}
\begin{equation}\label{lowvert2}
\langle\, ccu \,| g(u\bar{d})|\, ccd \,\rangle
\langle \chi_{MS}| \chi_{MS}\rangle=g\, .
\end{equation}

From Eqs.~\eqref{upvert}, \eqref{lowvert1} and \eqref{lowvert2}, the 
diagram in Fig.~\ref{mbint} can be written as
\begin{eqnarray}\label{amp}
-iT_{\Xi_{cc}\pi\to \Xi_{cc}\pi}&=&\sum\limits_{diagrams}(-i)V_{upper}\,iG_{prop}\,(-i)V_{lower}\nonumber\\
&=&\frac{2}{3}2ig\frac{1}{\sqrt{2}}(p_0+p^{\prime}_0)
\Big(\frac{-i}{-M^2_{\rho}}\Big) (i)\Big(\frac{-g}{\sqrt{2}}\Big)\nonumber\\
&+&2\sqrt{\frac{2}{3}}\sqrt{\frac{1}{3}}
(-i)g\frac{1}{\sqrt{2}}(p_0+p^{\prime}_0)\Big(\frac{-i}{-M^2_{\rho}}\Big)ig\, ,
\end{eqnarray}
and thus,
\begin{equation}\label{ampfinal}
T_{\Xi_{cc}\pi\to\Xi_{cc}\pi}=\Big(-\frac{4}{3}\Big)\frac{1}{4f^2_{\pi}}(p_0+p^{\prime}_0)\, .
\end{equation}
$V_{upper}$ and $V_{lower}$ in Eq.~\eqref{amp} stand for the upper and lower vertices, 
respectively, while $G_{prop}$ is the propagator related to the vector meson 
exchanged. In Eq.~\eqref{amp} we have used $g=M_V/2f_{\pi}$ to simplify the final 
expression, given by Eq.~\eqref{ampfinal}. Thus, the coefficient $C_{ij}$ for this transition is $-4/3$, as can be seen in Table~\ref{tabCij}. 
Analogously, all the other transition amplitudes can be obtained following the steps shown above.

%\newpage

\end{document}